\input head
\begin{document}
\begin{titlepage}
\vspace*{2.0cm}
\begin{center}
{\large\bf Gravitational microlensing by the halo of
the Andromeda galaxy}\\
\vskip 2.0cm
{\bf Philippe~Jetzer* }\\
\vskip 1.0cm
Institute of Theoretical Physics,
University of Z\"urich, Sch\"onberggasse 9,\\
CH-8001 Z\"urich, Switzerland
\end{center}

\vskip 2.0cm
Code numbers: 03.12.2, 07.30.1
\vskip 2.0cm
Section: 4 (Galactic structure and dynamics)
\vskip 2.0cm
Publication proposed in: Main Journal (Research Note)
\vfill
$^*$ Supported by the Swiss National Science Foundation.
\end{titlepage}

\newpage

\begin{center}
Abstract
\end{center}
\begin{quote}
It has been shown by Paczy\'nski that gravitational
microlensing is a useful method to detect brown dwarfs in the
dark halo of our galaxy.
At present several experiments are carried out to monitor
several million stars in the Large Magellanic Cloud
with the aim to find such microlensing events.

Here
I discuss the possibility to use as targets stars in the Andromeda
galaxy. Such an experiment would be sensitive to both the brown dwarfs
in our halo and in the one of M31.
The optical depth $\tau$ to gravitational microlensing due to brown
dwarfs in the halo of M31 turns out to be $\sim 10^{-6}$, which
is comparable
to the value of $\tau$ for our own halo.
I also compute
the microlensing rate and the average lensing duration and consider
moreover the dependence of $\tau$ on the shape of
the halo.
\end{quote}

\newpage
\noindent{\bf 1. INTRODUCTION}\\

Several observations, in particular the measurements of the rotation curves
(Faber and Gallagher 1979; Rubin
et al. 1980; Sancisi and van Albada 1987),
indicate that galaxies are embedded in a massive dark halo, which could
be made of compact baryonic objects (we shall call them ``Massive Halo
Objects'': MHO) such as brown dwarfs or jupiters
(Carr et al. 1984; Rees 1986; De R\'ujula et al. 1992).
Such objects are aggregates of H and He in proportions as predicted by
primordial nucleosynthesis.

Paczy\'nski 1986 has shown that gravitational microlensing is
a viable method to detect MHOs in the halo of our own galaxy
if their mass lies in the approximate range $10^{-6}
\leq M/M_{\odot} \leq 10^{-1}$.
A microlensing event occurs when a MHO crosses close to
the line of sight of a distant and ``fixed''
star.
The deflection of the star's light by the MHO's gravity
produces an (unresolvable) multiple image and an observable
time-dependent and colour-preserving
magnification of the original  brightness of the source.

Observations  which are now under way
(Ferlet et al. 1990; Moniez 1990; Griest et al. 1991)
monitor up to several million stars in
the LMC and
in a starfield nearby the Galactic Bulge
(Udalski et al. 1992).
In the meantime it has also been proposed to use as targets stars in
the Andromeda galaxy (Crotts 1992; Baillon et al. 1992).

The aim of this note is to study in some detail this latter possibility, which
would also allow to
detect MHOs in the halo of M31 and similarly in M33.
It turns out that the massive dark halo of M31 has an
optical depth to microlensing which is of about the same order of magnitude
as that
of our own galaxy $\sim 10^{-6}$, a fact this which remains also true
if the halo is flattened. This means that there is
a probability of $\sim 10^{-6}$ that a star im M31 is strongly
gravitationally microlensed by a MHO of his own halo.
Moreover, an experiment monitoring
stars in M31 would be sensitive to both the MHOs in our halo
and in the one of M31.\\

\noindent{\bf 2. OPTICAL DEPTH $\tau$}\\

We consider a flat space and a pointlike source as well as a pointlike
deflector. Then
the equation of gravitational lensing in the deflector's plane can be
written as
\begin{equation}
r^2-r_0 r-R_E^2=0~,  \label{eq:a}
\end{equation}
where the origin of the coordinate system is located on the lensing mass.
The source is at $r_0$ and the image at $r$. $R_E$ is
the Einstein radius defined as
\begin{equation}
R_E^2=\frac{4GMD}{c^2}x(1-x)~~~~0 \leq x \leq 1~, \label{eq:b}
\end{equation}
with $M$ the MHO mass and $D$ ($xD$) the distance from observer to source
(to the MHO). The magnification $A$ of the source due to lensing is given by
\begin{equation}
A=\frac{u^2+2}{u(u^2+4)^{1/2}}  \label{eq:c}
\end{equation}
with $u=r_0/R_E$.

The optical depth $\tau$ is equal to the ratio
of surface mass density of microlensing matter to the critical mass density,
which is defined as:
$c^2/(4\pi G D x(1-x))$ (see Vietri and Ostriker 1983; Nityananda and
Ostriker 1984 and Pacy\'nski 1986).
$\tau$ is the same as the probability that a source
is found within a radius $R_E$ around a MHO and gets therefore
magnified more than $A= 1.34$.
Since the mass density varies along the line of
sight, $\tau$ becomes the following integral
\begin{equation}
\tau=\int_0^1 dx \frac{4\pi G}{c^2}\rho(x) D^2 x(1-x)
\label{eq:d}
\end{equation}
with $\rho(x)$ the mass density of microlensing matter at distance
$xD$ from us. In our case $\rho(x)$ has two contributions, one from the
halo of our galaxy and a second one from the halo of M31.
This way the integral splits into two pieces.
The contribution arising from our own halo has been computed first
by Pacy\'nski 1986 using an isothermal spherical halo
model for the dark matter.
Griest 1991 generalized it by including
also a core radius $a$ (see also Jetzer 1991).

To estimate the contribution arising from the halo of M31
we consider for the mass a distribution that generalizes
the standard isothermal sphere, with
\begin{equation}
\rho(r)=\frac{\rho(0)}{1+(r/a)^2}~, \label{eq:e}
\end{equation}
where $\rho(0)$ is the density in the centre and $a$ is the core radius.
As an approximation we adopt for these quantities the same values
as for our
own galaxy. Typically $a \approx 5$ kpc and $\rho(0)=\rho_{\odot}(1+
(R_{GC}/a)^2)$. $R_{GC}$ is the distance of the Sun from the centre
of our galaxy and $\rho_{\odot}$ the density of the dark halo matter near
the Sun. $r$ is the distance of a MHO from the galactic centre of Andromeda.
$r=(D(1-x)+r_s)$, where $D$ is
the distance of Andromeda (or M33) from the Sun, which
is about 650 kpc (730 kpc). $r_s$ is the distance of the lensed star
from the galactic centre of M31, which we take in the range:
$0 \leq r_s \leq 5$ kpc. For simplicity we restrict ourselves to stars
located on the line of sight which goes through the galactic centre
of M31. This approximation should, however, be quite good, since only the
inner part of the
visible galactic disc with a radius of at most 5 kpc around the centre
contains enough stars for a microlensing experiment.
This is small compared to the extension of the halo, which can go as far
as $D_h$=50 kpc or even more.
Inserting eq.(\ref{eq:e}) into eq.(\ref{eq:d}) we get
\begin{equation}
\tau=\tau_0 \int^{1}_{x_h} \frac{x(1-x)}{\alpha^2+(1-x+\beta)^2}dx~,
\label{eq:f}
\end{equation}
\begin{equation}
\tau_0=\frac{4\pi G}{c^2} a^2 \rho(0) \approx 5 \times 10^{-7}~,
\label{eq:g}
\end{equation}
with $\alpha=a/D \approx 8 \times 10^{-3}$, $x_h=1-(D_h-r_s)/D \approx
0.92 - 0.93$ and $\beta=r_s/D \approx 0 - 8 \times 10^{-3}$ for
$0 \leq r_s \leq 5$ kpc.
The integral in eq.(\ref{eq:f}) can be evaluated analitically, which leads to
\begin{equation}
\eqalign{\frac{\tau}{\tau_0}= &
-\frac{(2\beta+1)}{2}ln~(\frac{\alpha^2+\beta^2}
{\alpha^2+(1-x_h+\beta)^2})+(x_h-1) \cr & +\frac{(\beta^2+\beta-\alpha^2)}
{\alpha}(arctan~(\frac{\beta}{\alpha})-
arctan(\frac{1-x_h+\beta}{\alpha})) \approx 1.7~.}  \label{eq:h}
\end{equation}
The numerical value is an average for $x_h$ and $\beta$
varying in the above mentioned ranges.
It turns out that typically
$\tau \sim 10^{-6}$ (similarly for M33).
This is about the same as one gets
for our own halo. Therefore, the total optical depth to microlensing
for a star in M31 is twice this value.

In the following we compute the optical depth $\tau$ for the case where the
halo is not spherical but rather axisymmetric, with an axis ratio
$cos~\psi$ (ratio of the z axis to the x or y axis). The generalization
of eq.(\ref{eq:e}) is
\begin{equation}
\rho(r)=\frac{\tan~\psi}{\psi}~ \frac{\rho(0)}{1+(\xi(r)/a)^2}~,
\label{eq:h1}
\end{equation}
where
\begin{equation}
\xi^2 \equiv r^2 + z^2tan^2\psi~, \label{eq:h2}
\end{equation}
and z is the symmetry axis (see Binney and Tremaine 1987; Gould 1993).
The disc of M31 has an angle of inclination $i \simeq 75^0$ (Braun 1991),
where $i$ is the angle between the line of sight to the Earth and the
normal (z axis) to the disc.
We consider line of sights which are at a distance b (impact parameter)
from the one which goes through the
centre of M31, and with a position angle $\phi$ relative to the
far side of the minor axis.
A line of sight with impact parameter
b (with $\phi=0$ or $=\pi$) hits the disc on the
minor axis at a distance $b/cos~i \simeq 3.86~b$
from the galactic centre. With $t=cos~i~tan~\psi$, $\xi^2$ of
eq.(\ref{eq:h2}) becomes
\begin{equation}
\xi^2= \frac{b^2}{cos^2\theta}+b^2 (tan~\theta+tan~i~cos~\phi)^2~ t^2~,
\label{eq:h3}
\end{equation}
where $\theta$ is the angle between the position vector
$\vec r$ (the location, with respect to the galactic centre of M31,
of a MHO on the line of sight)
and the line going through the
galactic centre and which is perpendicular to the line of sight. $\theta$ is
taken positive when the projection of $\vec r$ on the line of sight points
towards the observer, and negative otherwise.
With these definitions eq.(\ref{eq:d}) can be written as follows
\begin{equation}
\frac{\tau}{\tau_0}=\frac{tan~\psi}{\psi}~
\int^{\theta{max}}_{\theta_{min}} d\theta~
\frac{(tan~\theta+tan~i~cos~\phi)(1-\delta~tan~\theta-\delta~tan~i~cos~\phi)}
{(\frac{a}{b})^2 cos^2\theta+1+(tan~\theta+tan~i~cos~\phi)^2~t^2~cos^2\theta}
{}~, \label{eq:h4}
\end{equation}
where $\delta=b/D$.
For impact parameters on the far side of the minor axis ($\phi=0$)
$\theta_{min}=-i$, whereas for impact parameters on the near side of the
minor axis ($\phi=\pi$) $\theta_{min}=i$. $\theta_{max}$ is given using
the relation
$tan~\theta_{max}=\frac{\sqrt{\tilde D_h^2-b^2}}{b}$,
where $\tilde D_h$ is the extent of
the halo along the line of sight. For nonspherical halos we take
the extent of halo along the x or y axis to be $D_h=50~kpc$ and for the z axis
$D_h=cos~\psi \times 50~ kpc$.

In figure 1 the optical depth
$\tau/\tau_0$ is plotted as a function of the impact parameter
b along the minor axis
for a spherical halo and for two axisymmetric halos (corresponding
to $cos~\psi=0.75$ and $cos~\psi=0.5$).
For b=0 in the spherical halo case $\tau/\tau_0$ is given by eq.(\ref{eq:h})
by setting $\beta=0$ (or $r_s=0$). This way we get a slightly higher
value for $\tau/\tau_0~ (\simeq 2.2)$ as the one obtained in eq.(\ref{eq:h})
with $\beta \neq 0$.
\footnote{The value of $\tau$ for the spherical halo case given in the paper
by Crotts 1992 is higher with respect to ours by a factor of about 1.33,
due to the adopted value of $v_{rot}=280~km/s$ instead of
$v_{rot}=210~km/s$.}
{}From figure 1 we see that $\tau$ increases
by increasing the flatness of the halo. This is due to the fact that for M31
the z axis is almost perpendicular ($i \simeq 75^0$) to the line of sight
to the Earth. Thus with increasing flatness more brown dwarfs are located
along the line of sight.

$\tau$ increases by increasing the impact parameter b along
the far side of the minor axis, whereas decreases in the opposite direction.
This is due to the fact that the line of sight in the former case crosses
a larger portion of the denser inner part of the halo.
This depends on the favourable inclination of the disc of M31, a fact which
has been pointed out by Crotts 1992, and which may be used to
distinguish lensing events from variable stars by comparing the
microlensing rates for different starfields of the disc.\\

\noindent{\bf 3. MICROLENSING RATE}\\

The microlensing rate depends on the mass and the velocity distribution of
the MHOs. For a detailed derivation of the rate we refer to
De R\'ujula et al. 1991 and Griest 1991.
The mass density at a distance $xD$ from the observer is given by
eq.(\ref{eq:e}). The isothermal
spherical halo model does not determine the MHO number density as a
function of mass. A
simplifying  assumption is to let the mass distribution be independent
of the position in the galactic halo, i.e., to assume the following
factorized form for the number density per unit mass $dn/dM$
($\mu=M/M_{\odot}$),
\begin{equation}
\frac{dn}{dM}dM=\frac{dn}{d\mu}d\mu
=\frac{dn_0}{d\mu}H(x) d\mu~,\label{eq:zi}
\end{equation}
with
\begin{equation}
H(x)=\frac{a^2}{a^2+D^2(1-x+\beta)^2}~,
\label{eq:zj}
\end{equation}
and $n_0$ not depending on $x$.
For a nonspherical halo we get instead: $H(r)=\rho(r)/\rho(0)$,
with $\rho(r)$ given by eq.(\ref{eq:h1}).
It is $n_0$ that carries the
information on the MHO mass distribution and it is subject to the
normalization
\begin{equation}
\int d\mu \frac{dn_0}{d\mu}M=\rho(0)~.\label{eq:zl}
\end{equation}
Nothing a priori is known on the distribution $d n_0/dM$.

A different situation arises for the velocity
distribution, which for the isothermal
spherical halo model is known (Chandrasekhar 1942; Primack and Seckel 1988;
Krauss 1989). The projection of the velocity distribution
in the plane perpendicular to the line of sight leads to the following
distribution of the transverse velocity $v_T$
\begin{equation}
f(v_T)=\frac{2}{v_H^2}v_T e^{-v^2_T/v_H^2}~,\label{eq:zr}
\end{equation}
with $v_H \approx 210~km/s$ as a typical velocity dispersion
in the halo.

Using the above mass and velocity distributions,
the rate $d\Gamma$ at which (at fixed $\mu, u, x$ and $v_T$)
a single star is microlensed by a MHO
in the interval $d\mu du dv_T dx$ becomes
\begin{equation}
d\Gamma=2v_T f(v_T)D r_E [\mu x(1-x)]^{1/2} H(x)
\frac{d n_0}{d\mu}d\mu du dv_T dx~,\label{eq:zt}
\end{equation}
with
\begin{equation}
r_E^2=\frac{4GM_{\odot}D}{c^2} \sim (73~ a.u.)^2 \sim
(1.1 \times 10^{10} km)^2~ .\label{eq:zs}
\end{equation}
For an experiment, that monitors $N_{\star}$ stars during a total
observation time $t_{obs}$, the differential number of
microlensing events is
\begin{equation}
dN_{ev}=N_{\star} t_{obs} d\Gamma~,\label{eq:tf}
\end{equation}
$N_{ev}$ is obtained by integration
over an appropriate range for $\mu$, $x$,
$u$ and $v_T$, which takes into account the fact that an
experiment is sensitive to events with
a magnification such that $A \geq A_{min}$ and has time
thresholds $T_{min} \sim {\cal O}$(tens of minutes) and
$T_{max} \sim {\cal O}$(years)
(see De R\'ujula et al. 1991).

In order to quantify $N_{ev}$ it is convenient
to take as an example a delta function distribution for the mass.
For an experiment monitoring $N_{\star}$ stars during an
observation time $t_{obs}$ the total number of events with a
magnification $A \geq A_{min}$ (or equivalently $u \leq u_{max}$) is
\begin{equation}
N_{ev}(A_{min})=N_{\star} t_{obs} \Gamma(A_{min})~,  \label{eq:ll}
\end{equation}
with
\begin{equation}
\Gamma(A_{min}) =
D r_E u_{max} \sqrt{\pi}~v_H \frac{\rho(0)}
{M_{\odot} \sqrt{\bar \mu}}
\int^{1}_{x_{h}} dx[x(1-x)]^{1/2} H(x)
=\tilde \Gamma_0 u_{max}~.\label{eq:ta}
\end{equation}
By taking an average value for
the integral, which depends on $\beta$ and $x_h$, we get
for M31 ($D$=650~kpc)
\begin{equation}
\tilde\Gamma_0=1.8
\times 10^{-12}~\frac{1}{s}~\left( \frac{v_H}{210~km/s}\right)
 \left(\frac{1}{\sqrt{D/kpc}}\right)
 \left( \frac{\rho(0)}{1GeV/cm^3}\right)
\frac{1}{\sqrt{M/M_{\odot}}}\ .
\label{eq:tb}
\end{equation}
The average lensing time is given by
\begin{equation}
< T > \sim (125~ days)~ \sqrt{M/M_{\odot}}~. \label{eq:tc}
\end{equation}
In the following table we show some values of $N_{ev}^a$ due to MHOs in the
halo of M31 with
$t_{obs}=10^7$ sec ($\sim$ 4 Months), $N_{\star}=10^6$ stars and
$A_{min} = 1.34$ (which corresponds to an increase in
magnitude of $\Delta m = 0.32$). (We take $v_H=210~km/s$ and
$\rho(0)=1~Gev/cm^3$.) In the last column we give
the corresponding number of events due to MHOs in our own galaxy.
The mean microlensing time turns out to be about the same for both
types of events.

\begin{center}
\begin{tabular}{|c|c|c|c|c|}\hline
MHO mass in units of $M_{\odot}$ & Mean $R_E$ in km & Mean microlensing time &
$N_{ev}^a$ & $N_{ev}$ \\
\hline
&  &  &  & \\
$10^{-1}$ & $7 \times 10^8$ & 38 days & 2 & 1 \\
$10^{-2}$ & $2 \times 10^8$ & 12 days & 7 & 4 \\
$10^{-4}$ & $2 \times 10^7$ & 30 hours & 70 & 43 \\
$10^{-6}$ & $2 \times 10^6$ & 3 hours & 700 & 430 \\
\hline
\end{tabular}
\end{center}
\vskip 1.0cm

$N_{ev}^a$ is almost by
a factor of two bigger than
$N_{ev}$ (see also Baillon et al. 1992).
Of course these numbers should be taken as an estimate, since they
depend on the details of the  model
one adopts for the distribution of
the dark matter in the halo.
To distinguish between events due to a MHO in our halo or in the one of
M31 might be rather difficult.
Nevertheless, if microlensing
events are discovered by experiments monitoring the LMC,
it would then be possible to compute quite accurately $N_{ev}$ and,
therefore, to determine $N_{ev}^a$.

For a nonspherical halo $\Gamma$ is computed using $\rho(r)$ as given
by eq.(\ref{eq:h1}) and the notations mentioned in section 2.
We evaluated $\Gamma$ as a function of the impact parameter b
along the minor axis for
nonspherical axisymmetric halos with $cos~\psi=0.75$ and $cos~\psi=0.5$.
In figure 2~ $\gamma(b)=\Gamma(b)/\tilde\Gamma_0 u_{max}$ is plotted
for the two nonspherical halo models as well as for a spherical one.
For b=0 in the spherical halo case $\gamma$ is obtained by eq.(\ref{eq:ta})
with $\beta=0$, in which case we get $\gamma(0)=1.34$. As for
$\tau$ by increasing the flatness of the halo $\Gamma$ increases, whereas
the average lensing duration remains practically unchanged.\\

\noindent{\bf 4. CONCLUDING REMARKS}\\

An important fraction of the brightest stars in M31 are
red giants.
For these stars the apparent radius projected on the deflector's plane,
which is
$R_p=R_{star} x$ ($R_{star}$ being the radius of the star),
can be of the same size as the Einstein radius
$R_E$ of the lensing object. Therefore, the point-source approximation
will be no longer valid.
The magnification is large provided that $R_p \ll R_E$, whereas
it will be small if $R_p \approx R_E$. From this
condition one can estimate a minimum MHO mass below which the
magnification becomes negligible (see Paczy\'nski 1986):
\begin{equation}
\eqalign{M_{min} & \approx \frac{c^2}{4 G D x(1-x)} R_p^2
=\frac{c^2}{4 G D x(1-x)}(R_s x)^2 \cr &
\approx \frac{c^2}{4 G D_{ds}} R_s^2 \cr &
\approx~3 \times 10^{-7}~M_{\odot}~
\left( \frac{10~kpc}{D_{ds}}\right)
\left( \frac{R_{star}}{R_{\odot}} \right) ^2 ,} \label{eq:wnn}
\end{equation}
where $D_{ds}=D(1-x)$ is the distance between the MHO and the star. We
used the fact that $x \approx 1$.
{}From eq.(\ref{eq:wnn})
it follows that for a red giant with $R=100 R_{\odot}$ and
for a typical value $D_{ds}=20~ kpc$
the minimum lensing mass has to be
$\approx 10^{-3}~M_{\odot}$.
We thus expect microlensing to be sensitive to MHOs in the halo of M31, which
are in the mass
range: $10^{-3} \leq M/M_{\odot} \leq 10^{-1}$.
For MHOs in the halo of our own galaxy the corresponding range is:
$10^{-6} \leq M/M_{\odot} \leq 10^{-1}$. This has to be
compared with $10^{-4} \leq M/M_{\odot} \leq 10^{-1}$
when the targets are giant stars in the LMC.

To summarize, we showed that the optical depth to microlensing
of stars in M31 due to its halo is comparable to that of our
own galaxy. Due to the favourable inclination of the disc of M31
$\tau$ is increased if the halo turns out to be
flattened (along the z axis).
Should microlensing events be discovered,
by comparing the rates from experiments looking at the LMC and at M31,
one could extract informations on the dark halo of the Andromeda galaxy.
Finally, we mention that another way
to detect brown dwarfs in nearby galaxies has been suggested recently
by Daly and McLaughlin 1992. They pointed out that the next
generation of infrared satellites to be put in orbit like ISO or SIRTF will
be able to measure the integrated infrared emission of brown dwarfs
with a mass in the range $\sim 10^{-1}$ to $10^{-3} M_{\odot}$ located
in the halo of nearby galaxies.\\

I would like to thank A. Bouquet for useful suggestions.\\

\noindent{\bf REFERENCES}\\
\begin{itemize}
\item Baillon, P., Bouquet, A., Giraud-H\'eraud, Y., and Kaplan, J. 1992,
Preprint PAR-LPTHE 39.
\item Binney, J., and Tremaine, S., 1987, {\it Galactic Dynamics}
(Princeton Univ. Press).
\item Braun, R., 1991, Astrophys. J. {\bf 372} 54.
\item Carr, B.J., Bond, J.R., and Arnett, W.D., 1984, Astrophys. J.
{\bf 277}, 445.
\item Chandrasekhar, S. {\it Principles of Stellar Dynamics}
(The University of Chicago Press, Chicago 1942).
\item Crotts, A.P., 1992, Astrophys. J. {\bf 399}, L43.
\item Daly, R.A., and McLaughlin, G.C., 1992, Astrophys. J. {\bf 390},
423.
\item De R\'ujula, A., Jetzer, Ph. and Mass\'o, E., 1991,
Mont. Not. R. Astr. Soc. {\bf 250}, 348;
\item De R\'ujula, A., Jetzer, Ph. and Mass\'o, E., 1992,
Astron. \& Astrophys. {\bf 254}, 99.
\item Faber, S.~M. and Gallagher, J.~S., 1979, Ann. Rev. Astron.
Astrophys., {\bf 17}, 135.
\item Ferlet, R. {\it et al.}, 1990, Proposal to the C.A.S.E. Saclay.
\item Gould, A., 1993, Astrophys. J {\bf 404}, 451.
\item Griest, K., 1991, Astrophys. J {\bf 366}, 412.
\item Griest, K et al. 1991, Astrophys. J {\bf 372}, L79.
\item Jetzer, Ph., 1991, University of Z\"urich preprint ZU-TH 15, to appear
in ``Atti del Colloquio di Matematica'' Vol. {\bf 7}, Ed. CERFIM, Locarno.
\item Krauss, L. {\it The Fifth Essence: The Search for Dark Matter in
the Universe} (Basic Books. Inc. Publishers 1989).
\item Moniez, M., 1990, In Proceedings of the XXVth Rencontres de Moriond.
(Les Arcs, 1990) page 161.
\item Nityananda, R., and Ostriker, J.P. 1984, Ap. Astr. {\bf 5}, 235.
\item Paczy\'nski, B., 1986, Astrophys. J., {\bf 304}, 1.
\item Primack, J.~R. and Seckel, D., 1988,
Ann. Rev. Nucl. Part. Sci., {\bf 38}, 751.
\item Rees, M.J., 1986, in: {\it Baryonic dark matter}, 2nd ESO-CERN
Symposium.
\item Rubin, V.~C., Ford, W.~K. and Thonnard, N., 1980, Astrophys. J.,
{\bf 238}, 471.
\item Sancisi, R. and van Albada, T.~S.  in:
{\it Observational Cosmology}, IAU
Symposium 124 (Reidel, 1987) and in:
{\it Dark Matter in the Universe}, IAU
Symposium 117, Kormandy and Knapp eds. (Reidel, 1987).
\item Udalski, A., Szyma\'nski, M., Kaluzny, J., Kubiak, M. and Mateo, M.,
1992, Acta Astronomica {\bf 42}, 253.
\item Vietri, M., and Ostriker, J.P. 1983, Astrophys. J. {\bf 267}, 488.
\end{itemize}
\begin{figcap}
\item The optical depth $\tau/\tau_0$ as a function of the impact parameter b.
Positive values of b correspond to the far side of the disk
on the minor axis, whereas negative ones on the near side of the disk.
The continuous line is for a spherical halo. The dashed line is for
a nonspherical halo with $cos~\psi=0.75$ and the dotted line with
$cos~\phi=0.5$. We adopted for the core radius a value of $a \approx 5~kpc$.

\item The value of $\gamma=\Gamma/\tilde\Gamma_0 u_{max}$
as a function of the impact parameter b as in fig.1.
The meaning of the curves is the same as in fig.1.

\end{figcap}

\end{document}